\documentclass[bjps]{imsart}

\RequirePackage{amsthm,amsmath,amsfonts,amssymb}
\RequirePackage[authoryear]{natbib}
\RequirePackage[colorlinks,citecolor=blue,urlcolor=blue]{hyperref}
\RequirePackage{graphicx}
\startlocaldefs
\numberwithin{equation}{section}
\theoremstyle{plain}
\newtheorem{theorem}{Theorem}
\theoremstyle{remark}
\newtheorem{example}{Example}

\newcommand{\verythinmathskip}{\mskip 1mu}
\newcommand{\df}[1]{\emph{#1}}

\newcommand{\maj}{\mathrm{maj}}

 \newcommand\setof[1]{\mathopen\{\verythinmathskip#1\verythinmathskip\mathclose\}}

\newcommand\setOf[2]{\mathopen\{\verythinmathskip#1 : #2\verythinmathskip\mathclose\}}
\newcommand\Pbof[1]{\Prob\mathopen\{\verythinmathskip#1\verythinmathskip\mathclose\}}
\DeclareMathOperator{\Expv}{\mathsf{E}}
\newcommand\Expvof[1]{\Expv\mskip 1mu\mathopen\{\verythinmathskip#1\verythinmathskip\mathclose\}}

\newcommand{\clint}[2]{[#1,#2]}
\newcommand{\opint}[2]{(#1,#2)}

\newcommand \eps{\varepsilon}

\newcommand{\cE}{\mathcal{E}}
\newcommand{\bbR}{\mathbb{R}}
\newcommand{\bbS}{\mathbb{S}}
\newcommand{\bbZ}{\mathbb{Z}}

\newcommand{\sites}{{\Lambda}}
\newcommand{\states}{{\mathbb{S}}}
\newcommand{\nb}{u}
\newcommand{\Nbs}{U}

\newcommand{\trans}{g}
\newcommand{\Toom}{\text{Toom}}
\newcommand{\trpr}{\theta}
\newcommand{\flip}{F}
\newcommand{\refl}{R}
\newcommand{\GKL}{K}
\renewcommand\Pbof[1]{\mathrm{Prob}\mathopen\{\verythinmathskip#1\verythinmathskip\mathclose\}}
\newcommand{\zeros}{\mathbf{0}}
\newcommand{\ones}{\mathbf{1}}
\newcommand{\conv}{\mathrm{conv}}

\endlocaldefs

\begin{document}

\begin{frontmatter}

\title{Probabilistic cellular automata with Andrei Toom}
\runtitle{Probabilistic cellular automata with Andrei Toom}

\begin{aug}
\author[A]{\fnms{Peter}~\snm{G\'acs}\ead[label=e1]{gacs@bu.edu}}
\address[A]{Boston University\printead[presep={,\ }]{e1}}

\end{aug}

\begin{abstract}
  Andrei Toom, who died in September 2022, contributed some of the most
  fundamental results on probabilistic cellular automata.  We want to acquaint
  the reader with these and will also try to give the reader a look at the
  environment in which they were born.  Toom was an original and strong
  personality, and other aspects of his life (education, literature) will also
  deserve mention.
\end{abstract}


\end{frontmatter}

Andrei Toom, a key developer of the theory of probabilistic cellular automata, died in September 2022.
In this space I will describe his most important results, adding some evaluation of their significance.
In a closing section I will also refer to Andrei's life and the environment in which he acted.

\section{Probabilistic cellular automata}

Cellular automata are an attractive mathematical structure: simple to define, they give
rise to highly complex behavior.
They can model---in a qualitative way---a number of phenomena in physics, biology, society.
And they offer a number of natural, and at the same time very challenging, mathematical problems.
One way to think of them is as of a discrete generalization of partial differential equations.

A cellular automaton has a finite or countable number of units, its cells, typically arranged
on a finite-dimensional lattice, its set of \df{sites} \( \sites \);
they interact with their neighbors in a way that is uniform in space and time.
Here \( \sites \) will always be either the set \( \bbZ^{d} \) of points
in the \( d \)-dimensional space with integer coordinates, or a finite version of it, say
\( \bbZ_{m}^{d} \), where \( \bbZ_{m} \) is the set of remainders modulo \( m \).
Each cell has some \df{state}, belonging to some finite set \( \states \).
A \df{configuration} is a function \( \xi:\sites\to\states \), that is \( \xi(x) \) is the state
of the cell sitting at site \( x \).
The system develops in discrete time: our \df{space-time} is given by \( \sites\times\bbZ_{+} \).
A \df{history} is a function \( \eta: \sites\times\bbZ_{+}\to\states \), so \( \eta(x,t) \) is the state of
cell \( x \) at time \( t \).

The interaction between cells is local: the state of a cell at site \( x \) at time \( t+1 \) is only influenced
by the state of its \df{neighbors} \( x+\nb_{1},\dots,x+\nb_{n} \)  at time \( t \).
The list \( \nb_{1},\dots,\nb_{n} \) is the same for all \( x \).
In the case of finite space \( \sites=\bbZ_{m}^{d} \) the addition is taken modulo \( m \)
(this is called \df{periodic boundary conditions}).
Example for \( d=2 \): the list 
\begin{align}\label{eq:vonNeumannNbs}
(0,0),(-1,0),(1,0),(0,-1),(0,1) 
\end{align}
is called the \df{von Neumann neighborhood}.
The most important part of the definition of cellular automata is the kind of constraints
made on the history \( \eta(x,t) \).
We obtain a \df{deterministic} cellular automaton by specifying a \df{transition function}
\( \trans:\states^{n}\to\states \) and requiring that for each time \( t \),
each cell get its state at time \( t+1 \) from the state of its neighbors at time \( t \) as follows:
\begin{align}\label{eq:traj-req}
   \eta(x,t+1)=\trans(\eta(x+u_{1},t),\dots,\eta(x+u_{n},t)).
\end{align}
For example, for \( d=1 \) and the neighborhood \( \setof{-1,0,1} \) the requirement would be 
\begin{align*}
   \eta(x,t+1)=\trans(\eta(x-1,t),\eta(x,t),\eta(x+1,t)).
\end{align*}
Another way to express this is to say that there is a function, or ``operator'',
\( D:\states^{\sites}\to\states^{\sites} \)
taking configurations to configurations such that
\begin{align*}
   (D\xi)(x)=\trans(\xi(x+u_{1}),\dots,\xi(x+u_{n})).
\end{align*}
A history satisfying the requirement~\eqref{eq:traj-req} will be called a \df{trajectory} of the cellular automaton
having the transition function \( \trans \).
The initial configuration \( \eta(\cdot,0) \) and the transition \( D \) completely determine
the trajectory: \( \eta(\cdot,t)=D^{t}\eta(\cdot,0) \).

Apparently the first deterministic cellular automata were introduced by von Neumann and Ulam in the
1940's with the intention of modeling biological systems.
Von Neumann's explorations were cut short by his death, though their record is
available in~\cite{NeumannBurks66}.
As for the mathematical theory, 
soon after their introduction it has been understood that even 1-dimensional cellular automata can
simulate computations performed on any other formally defined computing device 
(for example Turing machines, see more in Section~\ref{sec:rel-comp}).
Therefore it is hard to find interesting mathematical questions about cellular automata that are not
undecidable.
(We will see that Andrei Toom succeeded in this.)

It is natural to generalize cellular automata by allowing the cells to make their transition
in a stochastic way.
Then \( \eta(x,t) \) becomes a random process.
The role of the transition function \( \trans \) is taken over by a set of \df{transition probabilities}
\( \trpr:\states^{n+1}\to\clint{0}{1} \).
The value \( \trpr(s\mid r_{1},\dots,r_{n}) \) shows the probability of transitioning into state \( s \)
at time \( t+1 \) provided the neighbors at time \( t \) have states \( r_{1},\dots,r_{n} \) at time \( t \).
Of course
\begin{align*}
   \sum_{s\in\states}\trpr(s\mid r_{1},\dots,r_{n})=1.
\end{align*}
\begin{example}\label{xpl:faulty-1d-majority}
For \( d=1 \), \( \states=\{0,1\} \),
the neighborhood \( \setof{-1,0,1} \) and some \( \eps\in\clint{0}{1} \)
let the transition be the following: first take the majority of the states of the three neighbors,
then with probability \( \eps \) change it to the opposite value.
Thus 
\begin{align*}
 \trpr(\maj(a,b,c)\mid a,b,c)=1-\eps .
\end{align*}
\end{example}
Call a \df{cylinder set} in the set of histories any set of the form
\begin{align*}
   \setOf{\eta}{\eta(x_{1},t_{1})=s_{1},\dots,\eta(x_{k},t_{k})=s_{k}}.
\end{align*}
The set of histories will be equipped with the discrete topology: the
cylinder sets form a basis of its open sets.
The same works for the set of configurations.
A \df{random history} is defined by a probability measure on the Borel sets of the histories.
If \( f \) is a measurable function over the probability space and \( \mu \) is a proability
measure then we will denote by \( \mu f \) the expected value (integral) of \( f \) by \( \mu \).

It is assumed that the cells make their choices independently, so the probability distribution
\( \Pbof{\cdot} \) satisfies for any \( t \), \( x_{1},\dots,x_{k} \):
\begin{equation}\label{eq:trans-prob}
\begin{aligned}
  \Pbof{&\eta(x_{1},t+1)=s_{1},\dots, \eta(x_{k},t+1)=s_{k}\mid \eta(\cdot,t'), t'\le t}
  \\  &= \prod_{i=1}^{k}\trpr(s_{i}\mid\eta(x_{i}+u_{1},t),\dots,\eta(x_{i}+u_{n},t)).
\end{aligned} 
\end{equation}
It follows from this definition that \( \eta \) is a Markov process.
Just like given a transition function of a deterministic cellular automaton
and an initial configuration \( \eta(\cdot,0) \) the trajectory is completely defined,
given the transition probabilities and a probability distribution over the initial
configurations  \( \eta(\cdot,0) \), the random process \( \eta \) is completely defined.
(The initial probability distribution can, in particular, be the special one, \( \delta_{\xi} \),
concentrated on a single configuration \( \xi \).)
In fact, the transition probabilities \( \trpr(\cdot\mid\cdot) \) define a linear operator \( P \)
on the set of measurable functions over \( \bbS^{\sites} \) as follows: for all \( t \),
\begin{align*}
   (P f) (\eta(\cdot,t)) = \Expvof{f(\eta(\cdot,t+1))\mid \eta(\cdot,t)}
\end{align*}
where \( \Expvof{\cdot\mid\cdot} \) is the conditional expected value.
Functions \( f \) of particular interest are 
\begin{align}\label{eq:e-x-a}
  e_{x,a}(\xi)=
  \begin{cases}
       1 & \text{if } \xi(x)=a
\\   0 & \text{otherwise}.
  \end{cases}
\end{align}
This also defines a linear operator \( \mu\mapsto P\mu \) on the 
space of probability measures over \( \states^{\sites} \) via \( (P\mu)f=\mu(Pf) \).
If we denote the probability distribution of \( \eta(\cdot,t) \) by \( \mu_{t} \), then
\(   \mu_{t+1}=P\mu_{t} \).
Clearly, the values \( P\delta_{\xi} \) over all configurations \( \xi \) define \( P \)
completely: without confusion we can write \( P\xi=P\delta_{\xi} \).
The transition operator \( D \) of a deterministic cellular automaton can be seen
as a special case via the following extension of its domain: \( D\delta_{\xi}=\delta_{D\xi} \).
The distribution \( \mu \) will be called \df{stationary} (or also, \df{invariant}) if \( P\mu=\mu \).
It is well-known and not difficult to prove that there is always at least one stationary distribution.

The study of probabilistic cellular automata was started in the early 1960's by a group of mathematicians
around Ilya Piatetsky-Shapiro.
Toom recalls the circumstances vividly in~\cite{ToomPiatetski09}.
The atmosphere of political thaw and wave of optimism, lasting just a few years,
spurned new scientific initiatives, partly by opening areas that were off-limits before, being considered
bourgeois science (like theoretical biology and ``cybernetics'').
In the department headed by I.~I.~Piatetski-Shapiro in I.~M.~Gelfand's
laboratory of applied mathematics at the Moscow State University, a diverse and dynamic group
of young scientists explored a variety of potential applications.
For the few mathematicians among then (a minority), it took a while to focus on
the kind of nontrivial questions where an answer with mathematical rigor
could be expected.
This is how the model of probabilistic cellular automata emerged;
soon it became clear that even the simplest examples and the simplest natural
questions about them posed worthy challenges.

In Moscow, two other groups lead by prominent mathematicians
worked on related problems of theoretical
statistical mechanics: those of Roland L.~Dobrushin and Yakov G.~Sinai.
There was a fertile interaction between these three groups.
A closely related model of cellular automata, where time is continuous,
has been mostly developed in the United States under the name of \df{interacting
particle systems},
and has the good fortune of a monograph~\cite{Liggett85} by Thomas
Liggett to refer to.
The development of the discrete-time and continuous-time models was largely
independent of each other, probably somewhat due to the limited possibilities of contact
between their researchers.
Andrei Toom participated in the writing of at least two useful introductions:
\cite{DiscrLocalMarkov90} and~\cite{ToomProblems95}.

One of those simplest natural questions
is whether the operator \( P \) defining some probabilistic
cellular automaton has more than one stationary distribution.
If the space \( \sites \) is finite then the Markov process we defined
is a finite-state Markov chain;
the theory of these is well-developed.
The operator \( P \) can  then be represented by a matrix.
If every state is reachable from every other state
by a sequence of positive-probability transitions then 
the chain has a single stationary distribution.
A related question is whether for every initial distribution \( \mu \) the sequence
\( P^{t}\mu \) converges to this stationary distribution.
The answer is yes if and only if an additional condition is satisfied:
that there is a \( t \) such that every element of the matrix \( P^{t} \)
is positive (this excludes the possibility of ``cycling'').
Such a Markov chain is called \df{ergodic}.
An informal way to express the meaning of ergodicity is to say that an ergodic
process eventually \emph{forgets everything} about its initial state.

The question is therefore new only for infinite cellular automata---here
the state space is uncountable.
Ergodicity is defined for this Markov process in the same way: requiring that
for every initial distribution \( \mu \) the sequence
\( P^{t}\mu \) converge to one and the same (stationary) distribution.
It is natural to choose the sense of convergence here to be that of \df{weak} convergence,
which in this case says that \( (P^{t}\mu)(C) \) converges for every
cylinder set \( C \).
The problem of giving sufficient criteria of ergodicity has been given attention at the
very beginning, also because it is related to the question of phase transition in the
models of equilibrium statistical mechanics developed by Dobrushin, Lanford and Ruelle.
One of these sufficient conditions can be found in~\cite{Vasershtein69}.

The first interesting question posed to the group was to find an example of a
non-trivial \emph{non-ergodic} probabilistic cellular automaton.
Kurdyumov showed in~\cite{Kurd80} (strengthened in~\cite{ToomUndec00})
that the ergodicity question for probabilistic
cellular automata is undecidable, even if only 0, 1, \( 1/2 \) are allowed as local transition probabilities.
Toom in~\cite{ToomUndecCont00} proved a similar result for continuous-time systems.

When speaking of non-ergodicity we generally expect at least
two stationary distributions.
I am not aware of any non-trivial example of an infinite probabilistic cellular
automaton \( P \) with only one stationary distribution but where this distribution
does not attract all \( P^{t}\mu \).
By ``non-trivial'' I mean a neighborhood larger than one and
no prohibited transitions.
In continuous time (interacting particle systems), in one dimension there is no such example,
as proved by Mountford in~\cite{Mountford95}.
A one-dimensional example with some prohibited transitions is given in~\cite{ChassaingMairess11},
and a three-dimensional example with only quasi-local interaction (so the neighborhood is infinite)
is given in~\cite{JahnelKuelske15}.

\section{The Stavskaya model}\label{sec:Stavskaya}

A candidate example~\cite{StavskayaPiatetski68} of non-ergodic cellular automata
emerged from computer experiments carried out by O.~N.~Stavskaya.
It is one-dimensional, with state space \( \states=\{0,1\} \).
I will describe it by switching the states from 0, 1 to 1, 0, in order to bring out
the analogy to the interacting particle system called the \df{contact process}.
There is a parameter \( \eps\in\clint{0}{1} \).
The neighborhood is \( \setof{0,1} \), and transition probabilities are
\begin{align}\label{eq:Stavskaya}
  \theta(1\mid a,b)=
  \begin{cases}
                  0 & \text{if } a+b=0,
    \\    1-\eps & \text{otherwise}.
  \end{cases}
\end{align}
In words: call a site \df{healthy} if its state is 0, and \df{sick} otherwise.
A sick site will be healed ``spontaneously'' with probability \( \eps \).
A healthy site can only become infected by a sick right neighbor; this will
happen with probability \( 1-\eps \): one can say that infection happens with
certainty but then spontaneous healing is applied to the site immediately.
Let \( \zeros \) be the configuration consisting of all 0's and \( \ones \) the
one consisting of all 1's.
The distribution \( \delta_{\zeros} \) is trivially stationary.
It should be clear that when \( \eps \) is close to 1 then \( P^{t}\mu \)
converges to \( \delta_{\zeros} \) for every \( \mu \).
Simulations suggested the conjecture that for small \( \eps \) there are also other
stationary distributions because they found that if \( \eta(\cdot,0)=\delta_{\ones} \)
then the values \( \Pbof{\eta(x,t)=0} \) appeared to be bounded by a constant \( c<1 \).
In this case one could say that the system exhibits a \df{phase transition}.
This conjecture was proved in 1968, by M.~A.~Shnirman in~\cite{Shnirman68} and
by Toom in~\cite{ToomOnStavskayaEng68}.

Toom's proof is simpler and more useful in the long run, as
it exploits an important observation: that 
the random space-time history of this model can be viewed as a percolation.
Indeed, consider a graph on the points of the space-time history \( \bbZ\times\bbZ_{+} \),
where edges are from each point \( (x,t) \) to \( (x,t+1) \) and \( (x-1,t+1) \).
A space-time point is \df{closed} if spontaneous healing happens there:
this occurs with probability \( \eps \) independently for each \( (x,t) \).
Otherwise it is \df{open}.
Suppose that we start from the initial configuration \( \eta(\cdot,0) \) in which
every site is sick.
Then \( \eta(x,t)=1 \) if and only if there is an open path from time 0 to \( (x,t) \).

The proof that for small \( \eps \) there is an upper bound to \( \Pbof{\eta(x,t)=0} \)
independent of \( x,t \) uses an argument that became called the ``Peierls argument'',
or the ``contour argument'', and it goes by the following steps.
\begin{enumerate}
\item Assuming an unfavorable event at space-time point \( (x,t) \) find a finite set \( B \)
  of space-time points in the past responsible for it.
  (In the case of this proof, the set of closed
  points on paths between \( (x,t) \) and time 0.)

\item Find some combinatorial structure \( \Gamma(\eta,x,t) \) 
  defining a nonempty subset \( B'(\Gamma) \subseteq B\).
  (In the case of this proof, a certain space-time path separating \( (x,t) \) from time 0: a ``contour''.)
\item Bound the number of possible
  structures \( \Gamma \) with \( |B'(\Gamma)|=k \) by \( c^{k} \) for some
constant \( c \).
\end{enumerate}
This proves \( \Pbof{\eta(x,t)=0}\le\sum_{k}(\eps c)^{k} \), which is small for small \( \eps \).
In a later paper~\cite{ToomNonMark72}, Toom proved that an invariant measure
different from \( \delta_{\ones} \) cannot be very simple (like a Markov chain).
See also~\cite{DepoorterMaes06} in this context.
Later work of Toom, as mentioned in Section~\ref{sec:proofs}, uses a 
much more complex structure \( \Gamma \).

The monotonicity properties
based on the partial order of measures introduced by Mityushin in~\cite{Mityushin70} and
the above result imply a ``phase transition'':
there is a critical value \( \eps^{*}\in\opint{0}{1} \) such that for \( \eps<\eps^{*} \) the
system is non-ergodic and for \( \eps>\eps^{*} \) it is ergodic.
The result raises another natural question: how many stationary measures are
there in the non-ergodic case?
Of course, the convex combination of any two stationary measures is also stationary,
so what we are really asking about is whether there are only two \df{extremal}
stationary measures.
Three papers have proved this, under some conditions.
Their methods are different, and each is instructive in its own right.
(One of the conditions is the \df{translation-invariance} of the stationary measures.
Formally, 
for a configuration \( \xi \) let us define the configuration  \( T_{v}\xi \) translated by
the vector \( v \) as follows: \( (T_{v}\xi)(x)=\xi(x-v) \).
The translation of a function \( f:\states^{\sites}\to\bbR \) is defined then by
\( (T_{v}f)(\xi)=f(T_{v}\xi) \), and a translation of a measure \( \mu \) by
\( (T_{v}\mu)f=\mu(T_{v}f) \).
The measure \( \mu \) is translation-invariant if \( T_{v}\mu=\mu \) for all \( v \).)

In 1970, Vasiliev in~\cite{VasilievCorrelation70} uses the technique of correlation equations
developed in statistical physics.
It requires \( \eps \) to be small.
In the same year, Vasershtein and Leontovich in~\cite{VasershteinLeontovichInvariant70} prove the
result for all \( \eps<\eps^{*} \), using an elegant algebraic representation,
for translation-invariant measures.
Finally Toom's proof in 1998 in~\cite{ToomPeriodic98}, requiring both small \( \eps \)
and translation-invariance, combines three-way coupling with a contour argument in
percolation.
Each of the three papers generalizes the model in a different way.

\section{Positive rates}

In the Stavskaya model the invariance of the measure \( \mu_{0} \) is guaranteed by
the requirement that in~\eqref{eq:Stavskaya} a healthy cell with healthy neighbors never
becomes sick: such a transition is \df{prohibited}.
For a number of years, the conjecture was considered that a probabilistic cellular automaton
with no prohibited transitions---that is when all local transition probabilities
\( \trpr(\cdot\mid\cdot) \) are positive---is always ergodic.
In the corresponding models in continuous time we would talk of positive \df{rates} in place
of local transition probabilities.
Toom's best-known result is a refutation of this conjecture: he provided a whole family of
non-ergodic probabilistic cellular automata with no prohibited transitions.

Let us call a probabilistic cellular automaton with all-positive local transition probabilities
\df{noisy}.
It is not unjustifed to view the goal of defining a  non-ergodic cellular automaton
that is also noisy as a goal of ``error-correction''.
An ergodic automaton would erase eventually all information about the initial configuration, while
a noisy non-ergodic one would preserve some of it, despite the presence of noise.
In fact all members of the family of examples defined by Toom can be viewed as
follows: a deterministic cellular automaton is given performing some local
error-correcting action, but then its action is ``perturbed'' by
changing each transition with some (small but positive) probability to something else.
A formal way of looking at this is the following.
Recall the definition~\eqref{eq:e-x-a}.
For some value \( \eps\in\clint{0}{1} \),
we say that transition operator \( N \) is an \( \eps \)-bounded \df{noise operator} if for all
\( x\in\sites \), 
\begin{align}\label{eq:noise-op}
   (N\delta_{\xi})e_{x,\xi(x)}>1-\eps.
\end{align}
In words, it changes the value \( \xi(x) \) only with probability less than \( \eps \).
If \( D \) is our deterministic
error-correcting operation then the full transition would be \( ND \):
the action of \( D \) is ``perturbed'' by the noise \( N \).
What would be some candidate actions \( D \) for local error-correction?
Let our automaton have just two local states, \( \states=\{0,1\} \).
We would like to see a noisy automaton with the property that, for example,
for both \( j\in\{0,1\} \), 
if \( \eta(x,0)=j \) for all \( x \) then \( \Pbof{\eta(x,t)\ne j}<1/3 \) for all \( x,t \).
In one dimension let the neighbors be \( -1,0,1 \), 
and in two dimensions take the neighborhood~\eqref{eq:vonNeumannNbs}.
A good candidate error-correcting action seems to be taking the majority of all 
neighbor states.

In one dimension, this is doomed to failure.
A rigorous (complex) proof of ergodicity is given in 1987 by Gray in~\cite{Gray87} (for all monotonic two-state
nearest-neighbor rules in one dimension).
For continuous time, Gray gave in 1982 a simpler but still non-trivial proof in~\cite{Gray82}.
Here is an informal argument.
In a cellular automaton with \( 0\in\states \), in some configuration \( \xi \)
let us call an \df{island} a finite set \( S \) such that \( \xi(x)=0 \) if and only if
\( x\not\in S \).
Let \( j=0 \), so we are starting from a ``sea of zeros''.
The noise will, occasionally, create a large (say, size 10) island.
The local majority vote rule will not be too helpful in eliminating the island.
Indeed, it cannot do much about the ends.
Noise will make them fluctuate, essentially perform a random walk.
The size of the island, the difference of these random walks, is also a random walk.
It will eventually return to 0, but only in infinite expected time; in the meantime many
other islands arise.

In two dimensions the situation is better, but not ideal.
In the absence of noise the majority rule over the von Neumann
neighborhood~\eqref{eq:vonNeumannNbs}
would not shrink an island, say a square of size \( n \), at all.
Simulations (and informal arguments) show that in \df{unbiased}
noise it will be eliminated in about \( O(n^{2})  \) steps.
To define bias precisely, we introduce the flipping operation \( \flip \) for a function \( f \)
over \( \{0,1\}^{\sites} \) as follows: 
\begin{align}\label{eq:flip}
 (\flip f)(\xi) = f(1-\xi) .
\end{align}
We say that the noise operator \( N \) is \df{unbiased} with respect to flips
if \( N\flip=\flip N \).

It is not known whether the majority rule over the von Neumann neighborhood,
perturbed in a small but unbiased noise, can be ergodic.
But there are some special non-ergodic noisy versions, see for
example~\cite{KozlovVasiliev80}, and the more elaborate~\cite{LebMaesSpeer90}.
Let \( \states=\setof{-1,1} \).
With \( U=\setof{(0,1),(0,-1),(1,0),(-1,0)} \),
let us define the process \( P \) by the following relation:
If \( s \) is the sum of \( \eta(x,t) \) over the four neighbors \( \setOf{x+v}{v\in U} \)
of \( x \) then
\begin{align*}
 \Pbof{\eta(x,t+1)=-1}=\eps^{s}\Pbof{\eta(x,t+1)=1}.
\end{align*}
So for example if there are 3 neighbors with \( +1 \) and 1 neighbor with \( -1 \)
then the next state
is \( \eps^{-2} \) times more likely to become \( +1 \) than \( -1 \).
For \( j=0,1 \) let for \( x=(x_{1},x_{2}) \):
\begin{align*}
  \sites_{j}=\setOf{(x,t)\in\sites\times\bbZ_{+}}{x_{1}+x_{2}+t\equiv j\pmod 2}.
\end{align*}
We turn \( \sites_{j} \) into a graph, connecting \( (x,t) \) and \( (y,t+1) \) by an edge
whenever \( x \) and \( y \) differ by 1 in just one coordinate.
Then the process \( \setOf{\eta(x,t)}{(x,t)\in \sites_{0}} \) is a Gibbs state of the Ising model of
equilibrium statistical mechanics (of course the same holds for \( \sites_{1} \)).
It is known that for small \( \eps \) (corresponding to ``low temperature'')
the Ising model has more than one Gibbs state,
which makes for several invariant distributions for the process \( P \).
(In the continuous-time models called interacting particle systems,
the corresponding model is called the \df{stochastic Ising model},  see~\cite{Liggett85}.)
This is a very delicate process, though.
If the transition probabilities are changed ever so slightly to prefer the 1's over the \( -1 \)'s,
the process becomes ergodic.
In general, if the majority rule over the von Neumann neighborhood is perturbed
in a way that is, say, biased towards the 1's, then a large island will \emph{grow}!
The \( 1/n \) speed of shrinking provided (on average) by the  majority rule
is overpowered by the constant speed of growth.

\begin{figure}
  \[
    \fboxsep=0pt
    \fbox{\includegraphics[scale=0.25]{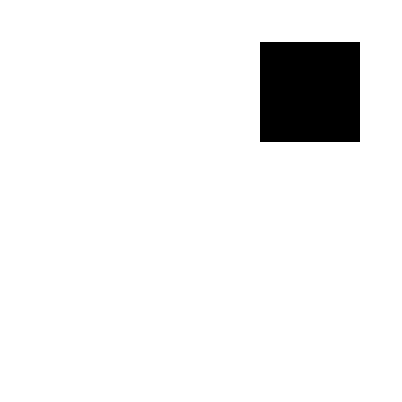}}\quad
    \fbox{\includegraphics[scale=0.25]{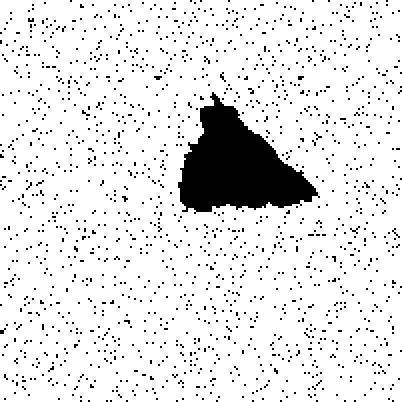}}\quad
    \fbox{\includegraphics[scale=0.25]{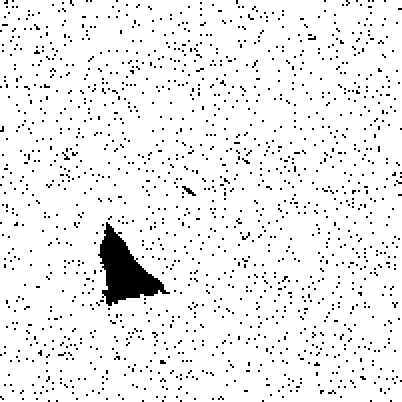}}\quad
    \]
  \caption{An island shrinking under Toom's rule (simulation).  The faults have probability \( 0.04 \) and are
    completely biased towards the 1's.}
\label{fig:Toom-shrinking-island}  
\end{figure}

Let us now turn to Toom's best-known example, in two dimensions, called the \df{Toom rule}:
\begin{align*}
   (D_{\Toom}\xi)(0,0)=\maj(\xi(0,0),\xi(0,1),\xi(1,0)).
\end{align*}
The transition rule is a majority vote over three neighbors: north, east,
self---which is then perturbed with some small probability \( \eps \).
The novelty is just that the neighborhood over which the majority is taken is
\emph{not (central-) symmetric}---but this makes a big difference.
Imagine an island.
Enclose it into a triangle with vertices \( (a,b),(a+c,b),(a,b+c) \).
If \( c\ge 1 \) then without noise, one application of the Toom rule will squeeze the island
into the triangle \( (a,b),(a+c-1,b),(a,b+c-1) \).
Thus, the island will shrink with \emph{constant speed}.
Small noise \emph{(even if it is biased!)} can slow down this shrinking,
but still leaves its speed constant (see Figure~\ref{fig:Toom-shrinking-island}).
 This last argument is not a proof; we will say more about the (not easy) proof in
the next section.

How small must be the probability \( \eps \) for non-ergodicity?
The existing proofs don't give an explicit bound, though for example (a rather bad) one can easily
be computed from the version of the proof in~\cite{GacsToom95}.
Simulations suggest that the upper bound \( 0.06 \) is sufficient.

\section{Reliable computation}\label{sec:rel-comp}

We pointed above to a connection of non-ergodicity with error-correction.
In its original form, non-ergodicity is only asking to safeguard at least one bit of
information---a minimal form of information conservation in the presence of local faults.
But the solutions found have much wider application.
Von Neumann in~\cite{VonNeum56} addressed the question of reliable computation with unreliable components.
His computation model is now called a Boolean circuit (for example with logic gates AND and NOT),
computing a Boolean function with, say, a one-bit output.
For a sufficiently small \( \eps \), for any Boolean circuit \( C \) of size \( N \)
he constructed another one, \( C' \), of size \( O(N\log N) \)
that computes the same output as \( C \) with probability \( 1-O(\eps) \)
even though each gate of \( C' \) is allowed to malfunction with probability \( <\eps \) (independently from
the others).
The construction multiplies each wire and each gate some \( O(\log N) \) times: in the absence
of faults, each wire in a bundle carries the same bit.
The key addition is to insert into each such wire bundle a little circuit called
the \df{restoring organ}.
The role of this organ is that if, say, the fraction of its input wires carrying faulty information
is less than \( 5\eps \) then after its application it will be reduced to less than \( 2\eps \): so it is about
protecting one bit of information!
Von Neumann used random permutations for the restoring organ; for constructive
(but still not local) solutions, see the survey~\cite{PippNeum88}.

As a model of computations, \df{Turing machines} are in several ways better than Boolean circuits.
They were introduced in the 1930's to formalize the theory of computability.
We define such a machine as follows.
It has some fixed finite set of possible internal \df{states}, and
a doubly infinite \df{tape} with symbols on it from a finite \df{alphabet}.
A read-write \df{head} is observing some tape position.
At each step, the machine performs an action based on a fixed \df{transition function}: depending
on the state and the observed tape symbol, it changes the state and the tape symbol, and moves the head
a step left or right.
The \df{input} is the initial content of the tape, and if at some time the machine
reaches a certain state called \df{halting state} then the \df{output} is the content found
at that time.
It is so widely accepted that every algorithm (in the informal sense) can be implemented
on an appropriate Turing machine that by now in mathematics a function (mapping inputs to outputs)
is called \df{computable} if it is computable on some Turing machine. 
There is a \df{universal} Turing machine, one that can simulate every other Turing machine.

Every Turing machine (in particular also the universal ones)
can be simulated by an appropriate one-dimensional
cellular automaton, so cellular automata can also serve well as
a model of arbitrary computation.
When asking for a reliable computer, it makes sense therefore to ask for a (noisy) probabilistic cellular
automaton capable of performing arbitrary computations (defined by its input).
Our 1988 construction with John Reif in~\cite{GacsReif3dim88}, built on Toom's rule, does this.
It takes an arbitrary (deterministic) one-dimensional cellular automaton \( A \) (say, a universal one),
say with neighborhood \( \{-1,0,1\} \), and
builds a \emph{three-dimensional} noisy cellular automaton \( A' \) simulating \( A \).
Let \( \zeta(x,t) \) be a trajectory of \( A \).
The intended history \( \eta(x,y,z,t) \) of \( A' \) would be
\( \eta(x,y,z,t)=\zeta(x,t) \): so ideally in \( A' \)
each cell of the whole plane \( \{x\}\times\bbZ^{2} \)
has the same state as the symbol at position \( x \) of \( A \).
If \( D_{A} \) is the transition rule of \( A \) then the rule of \( A' \)
says: in order to obtain your state at time
\( t+1 \), first apply the Toom rule in each plane defined by fixing the first
coordinate---call this \( D_{\Toom,2,3} \).
Then, apply rule of \( A \) on each line obtained by fixing the second and third
coordinates---call this \( D_{A,1} \).
Finally, apply an \( \eps \)-bounded noise operator \( N \), so the 
complete transition of \( A' \) is \( N D_{A,1}D_{\Toom,2,3} \).

The reliability of the automaton \( A' \) is proved in essentially the same way as the non-ergodicity
of the (perturbed) Toom rule; see Section~\ref{sec:proofs} for remarks on the proof.

It seems unrealistic for the automaton \( A' \) to store each symbol of information \( \zeta(x,t) \) 
with \emph{infinite redundancy}, as the array of all values \( \eta(x,y,t) \).
But if it is known that the computation of automaton \( A \) uses only \( S \) cells and
runs for only \( T \) steps then it can be run on the space \( \sites=\bbZ_{S} \), and can be
simulated reliably (failing only with probability \( O(\eps) \))
on the space \( \sites'= \bbZ_{N}\times\bbZ^{2}_{R} \) where
\( R = O(\log (ST)) \): so repeating each symbol only \( O(\log^{2}(ST)) \) times.
See the paper~\cite{BermSim88} by Berman and Simon, which improves on~\cite{GacsReif3dim88}.

\section{Eroders}

Toom called a deterministic cellular automaton an \df{eroder} if it eliminates any island.
From now on we will tacitly require \( \states=\{0,1\} \) and that the rule be \df{monotonic}.
The Toom rule \( D_{\Toom} \) leads to non-ergodicity because both it and its \df{dual} are
eroders, where the dual \( \flip D_{\Toom}\flip \) is defined using~\eqref{eq:flip}
(it happens that \( D_{\Toom} \) is self-dual).

What rules are eroders?
Mityushin and Toom in~\cite{MityushinToom74} proved that it is undecidable
about an arbitrary one-dimensional
monotonic binary cellular automaton whether it will erase the island \( \dots001100\dots \).
Mityushin and Toom also
define there a one-dimensional monotonic binary cellular automaton \( A \) for which
given an island it is undecidable whether \( A \) will erase it.

These statements make us appreciate Toom's elegant formula
of~\cite{ToomMon76} given in Theorem~\ref{thm:eroders} below---in any dimension \( d \)---that
decides about an arbitrary monotonic binary cellular automaton \( A \) whether it is an eroder.
Let \( \sites_{A}=\bbZ^{d} \), with neighborhood \( \Nbs_{A}\subset\sites \),
and transition rule \( \trans_{A}:\states^{\Nbs_{A}}\to\states \).
Recall that \( \states=\{0,1\} \).
We call a subset \( S\subseteq\Nbs_{A} \) a \df{null set} of \( A \)
if whenever in a configuration \( \xi \)
we have \( \xi(x)=0 \) for all \( x\in S \), this implies \( (\trans_{A}\xi)(0)=0 \).
For example, in the Toom rule, the minimal
null subsets are \( \{(0,0),(0,1)\} \), \( \{(0,0),(1,0)\} \), \( \{(1,0),(1,1)\} \).

\begin{theorem}\label{thm:eroders}
For a given cellular automaton \( A \) let
\begin{align*}
   \sigma(A)=\bigcap_{\text{null sets } S}\conv(S)
\end{align*}
where \( \conv(S) \) is the convex hull taken in the Euclidean space \( \bbR^{d} \).
Then \( A \) is an eroder if and only if \( \sigma(A) \) is empty.  
\end{theorem}
\begin{sloppypar}
It is important here that the the operation above happens in \( \bbR^{d} \).
Consider for example the automaton \( A \) in two dimensions with the transition 
\begin{align*}
  (D_{A}\xi)(0,0)=(\xi(0,0)\lor \xi(1,1))\land(\xi(0,1)\lor \xi(1,0)).
\end{align*}
The minimal nullsets are \( \{(0,0),(1,1)\} \) and \( \{(0,1),(1,0)\} \),
so \( \sigma(A)=\{(1/2,1/2)\} \), which is not in \( \bbZ^{2} \).
Still, by Toom's theorem, this is not an eroder.  
\end{sloppypar}

How about one dimension?
There are monotonic binary eroders in one dimension as well; however, none whose dual would also
be an eroder.

By the theorem of Toom given below, an eroder remains an eroder also in small noise.
It turns out that for this theorem we don't even need space-time uniformity or a
Markov property of the noise, so we can relax the requirement~\eqref{eq:noise-op} further,
using a formulation borrowed from~\cite{Toom80}.
Let \( A \) be an arbitrary deterministic cellular automaton, and \( \eta(x,t) \) a
random history of \( A \).
Let the random \df{fault set} \( \cE\subseteq\sites\times\bbZ_{+} \) be the subset of the space-time
in which \( \eta \) violates the transition rule \( \trans_{A} \).
We will say that the distribution of \( \eta \) is an \( \eps \)-\df{perturbation} of
\( A \) if for every finite subset \( S \) of space-time,
\begin{align}\label{eq:epsilon-perturbation}
   \Pbof{S\subseteq \cE}\le\eps^{|S|}.
\end{align}
A trajectory \( \zeta(x,t) \) of automaton \( A \) is called \df{stable} if there
is a function \( h(\eps) \) with \( h(\eps)\to 0 \) as \( \eps\to 0 \) such that
for all \( \eps \)-perturbations \( \eta \) of \( A \) starting from
the same initial configuration: \( \eta(\cdot,0)=\zeta(\cdot, 0) \), and for all \( x,t \)
we have \( \Pbof{\eta(x,t)\ne\zeta(x,t)}<h(\eps) \).
Call an eroder \df{stable} if the history \( \zeta \) with \( \zeta(x,t)=0 \) for all \( x,t \)
is a stable trajectory for it.

\begin{theorem}\label{thm:stable}
  Every binary eroder is stable.
\end{theorem}
As mentioned above, this theorem implies that the perturbed Toom rule is non-ergodic.

In continuous time, a simple characterization of eroders similar to Theorem~\ref{thm:eroders}
is not available.
But Gray gave a sufficient condition, and
proved in~\cite{Gray99} that transition rates corresponding to the Toom rule with a small noise
in continuous time are non-ergodic.
This result is also robust with respect to the kind of faults allowed: they can be biased.

\section{On the proofs}\label{sec:proofs}

Toom proved Theorem~\ref{thm:stable} using a kind of ``contour argument'' as outlined
in Section~\ref{sec:Stavskaya}.
The paper~\cite{Toom80} contains detailed proofs of both
theorems~\ref{thm:eroders} and~\ref{thm:stable}, in a somewhat more general setting.
In the contour argument the appropriate structure \( \Gamma(\eta,x,t) \) 
is quite complex: it is called a \df{truss} there.
A somewhat simplified version of this proof of just the stability of Toom's rule,
following~\cite{BermSim88}, can be found in~\cite{GacsToom95}, where the corresponding
structure is called an \df{explanation tree}.

Toom's first examples of noisy non-ergodic cellular automata
in~\cite{ToomNonerg73} used some special two-dimensional
eroders other than the north-east-center voting; their dual is also an eroder.
The contour argument in the proof of their stability is simpler, the structure \( \Gamma \) in
question is indeed only a kind of one-dimensional contour.
Here is one of these eroders, call it \( A \), with transition operator
\begin{align*}
  (D_{A}\xi)(0,0)=(\xi(0,0)\lor \xi(1,0))\land(\xi(1,0)\lor \xi(1,1)).
\end{align*}
Given any island, this rule squeezes it in every step into a narrower and
narrower horizontal stripe, eventually erasing it.
Its dual will squeeze an island in every step into a narrower and
narrower vertical stripe.

Our paper~\cite{GacsReif3dim88} contains a completely different kind of proof, based on a
hierarchical structure discoverable in the set of independent faults.
It is in some ways more intuitive as it is based directly on the picture
of shrinking triangles (see Figure~\ref{fig:Toom-shrinking-island}),
but leads to somewhat worse estimates.
The paper~\cite{BramsonGray91} by Bramson and Gray develops this method more systematically, introducing
a hierarchy of random processes built up from each other by a ``decoding'' operation.
Its method is also used in the proof of the continuous-time ``Toom theorem'' in~\cite{Gray99}.

\section{One dimension}

With Kurdyumov and Levin we defined a simple one-dimensional (not monotonic) two-state eroder (call it \( \GKL \))
in~\cite{GacsKurdLevIslands78} (it is sometimes referred to as the GKL rule).
Its dual is also an eroder; more precisely, recall the flip operation \( \flip \) in~\eqref{eq:flip},
and define the \df{reflection} operation \( \refl \) defined as \( (\refl\xi)(x)=\xi(-x) \) which
commutes with flip.
Then \( \GKL\flip\refl=\refl\flip\GKL \).
Alas, this rule (along with other simple ones similar to it)
is not stable, at least not in strongly biased noise, as shown
in Kihong Park's thesis~\cite{ParkThes96}.

As said above, there are no monotonic eroders in one dimension whose dual is also an eroder.
More generally, as also said, monotonic two-state one-dimensional
probabilistic cellular automata are ergodic: as proved by Lawrence Gray for continuous
time in~\cite{Gray82}, and for discrete time  in~\cite{Gray87}.
Based on this and also on the fact that in one-dimensional lattice equilibrium
systems (like the Ising model) there is no phase transition, and that therefore
the corresponding reversible Markov processes are ergodic,
the probabilistic cellular automata community formulated a hypothesis called
the ``positive rate conjecture'', saying that all noisy one-dimensional cellular automata
are ergodic.

Georgii Kurdyumov outlined in~\cite{Kurd78} a one-dimensional
cellular automaton that should be non-ergodic.
Its cells were meant to implement a hierarchical structure dealing with larger and larger
groups of faults.
However, details of the proposal did not follow.
I worked these out in~\cite{Gacs1dim86}, defining a cellular automaton that simulates
a similar one with some built-in error-correction;
this gives rise to an infinite hierarchy of more and more reliable cellular automata
all of which but the first one ``live'' in simulation.
A much more structured---and longer---paper~\cite{GacsSorg01} extended the result also to continuous time.
Its method uses the idea of Bramson and Gray in~\cite{BramsonGray91} of a hierarchy of random processes derived from
each other by a ``decoding'' operation.

These systems refute the positive rate conjecture.
As they are not monotonic and the corresponding Markov processes are not reversible either,
they don't contradict the motivating examples of the conjecture.

The one-dimensional constructions and proofs are very complex;
the cells have a very large number of states.
(One can say that as in one dimension the geometry does not help, all error-correction
must rely on ``organization''.)
The challenge to find simpler examples is still standing.

\section{Multi-level eroders}

After the beautiful characterization of two-state monotonic eroders in Theorem~\ref{thm:eroders},
Toom asked the question whether a similar characterization exists also when the set of states is
a finite ordered set, say \( \setof{0,1,\dots,n} \).
The situation turns out to be more complicated.
Galperin characterized in~\cite{GalperinEroder76} the one-dimensional eroders
in terms of the running speeds of the ends of islands of various levels.
But Toom showed in~\cite{ToomUnstable76} that some of these eroders,
even with just three levels,
are not stable, so the analogue of Theorem~\ref{thm:stable} does not hold. 
With Ilkka T\"orm\"a in~\cite{GacsTorma18} we characterized
all \emph{stable} one-dimensional multilevel eroders.

The question of characterizing multi-level eroders is still open in dimensions greater than one.
In one dimension even an unstable eroder erases islands in linear time.
In two dimensions, this is not always true.
The paper~\cite{MenezesToomNonlinear06} of Toom with his student
Menezes gives an example of a two-dimensional
three-level eroder that erodes some islands only in quadratic time.
In one-sided noise, it becomes ergodic.
This example can easily be modified to an eroder that erases some islands only in exponential time.
At this point it is not known whether
the question of which three-level monotonic two-dimensional cellular automata are eroders is decidable
at all.
The paper~\cite{SantanaRamosToom3states15} of Toom with some other students
analyses a related model in the Euclidean plane (instead of the grid).

Our method used in~\cite{GacsTorma18} seems generalizable to several dimensions,
using Toom's substantial sharpening in~\cite{ToomEstimates82} of his main stability result.
Therefore it is likely decidable about a monotonic multi-level cellular automaton whether it
is a stable eroder, while it might remain undecidable whether it is just an eroder.

\section{Other work on cellular automata}

Toom was one of the leading figures in Russian research on biologically and physically inspired systems
with local interaction throughout the 1970's and 80's.
He was one of the organizers of a regular conference in the biological research center in Pushchino,
and one of the editors, mostly along with R.~L.~Dobrushin and V.~I.~Kryukov, of the 
proceedings.
Unfortunately, most of these are in Russian, though some resulted also in an English-language
publication: see Selecta Mathematica Sovietica and~\cite{LocallyInteracting76}.
The book~\cite{DiscrLocalMarkov90} is an important and hard-to-access survey of research on various
aspects of probabilistic cellular automata.

The account in the present section of Toom's later work is incomplete, reflecting my own interests.
The technique developed for proving the main stable eroder theorem appears in several later publications.
The work~\cite{ToomReal78}, generalizes the cellular automata model in an unexpected direction:
space-time is a multi-dimensional Euclidean space.
However, the set of states is just \( \setof{0,1} \), so a history is just a subset of space-time.
Via the definition of a ``monotonic evolution'', the space-time set grows still in discrete
steps---allowing a generalization of the original eroder question.
Though the results are somewhat similar,
there is a richer set of possibilities: some eroders will erode only in non-linear time.
Systems that erode in linear time are characterized in a way similar to Theorem~\ref{thm:eroders},
via a set \( \sigma \); in the present case of a space-time system, instead of being empty, \( \sigma \)
has to consist of just the origin \( 0 \).
Now there will be some eroders even with \( \sigma\ne\{0\} \), just not working in linear time.
Noise is introduced and it is shown again that a linear-time eroder is also stable,
and if it is not eroding then it is not.
But only Toom's much later paper~\cite{Toom2002}
proves that the system is not a stable eroder when \( \sigma\ne\{0\} \).

The papers~\cite{ToomGrowthI94,ToomGrowthII94} strengthen the stable eroder results to cases
where the set \( \states \) of local states is not finite, but is the set of integers.
Of course, new conditions are needed on the transition probabilities.

The paper~\cite{ToomProhib79} places some cellular automata results into the broader
context of \df{tiling systems}, thus touching the area of symbolic dynamics.
Generally tiling questions don't involve probability,
but here a version of probabilistic perturbation with a condition similar to~\eqref{eq:epsilon-perturbation}
is introduced, so that questions of stability can be examined, allowing an application of the stable
eroder technique (among others).
Similar conditions on random perturbations have much later
been used by Durand, Romashchenko and Shen in~\cite{DurandRomashShenTiling12},
although for non-periodic tile sets, and a very different technique of ``error-correction''.
The perturbation condition raises a new challenge still mostly unsolved:
to find Gibbs distributions satisfying it.
The paper gives a construction only in one dimension, but even this with a quite technical proof.

The problem of information conservation
in one dimension continued to occupy Andrei Toom throughout
his career; he was not satisfied with the overly complex constructions
in~\cite{Gacs1dim86,GacsSorg01}.
In two papers, he explored alternative models: in these, the cells have a continuous
set of states (real numbers or real numbers modulo some \( M>0 \)).
The local transition rule is a (linear) averaging operation over the
neighborhood---with the intent of preserving a kind of continuity---plus some noise.
(There is reference to a related idea of Hammersley's ``harnesses''
in~\cite{HammersleyHarnesses67}.)
The paper~\cite{ToomTails97} gives a very detailed analysis of how the loss of continuity
depends on the tail behavior of the noise variables.

One idea for a finite system, also favored by physicists, is to store
the information in some kind of non-local,
topological characteristic, like rotation number.
In~\cite{ToomSuperexp95}, Toom explores this idea via a cellular automaton with a
finite number \( L \) of cells, periodic boundary conditions, whose local state space is 
the set of real numbers modulo \( M \).
Given the finiteness of \( L \), all information, including the rotation number,
will eventually be lost; the main
result of the paper lower-bounds the expected time for the loss of the rotation number
by approximately \( L^{-1}\exp(c M^{3}) \) (with a similar upper bound in some cases,
without the \( L \)).
So the information is preserved for a time growing
``super-exponentially'' as a function of the size \( M \) of the local state space.
The dependence on the size \( L \) of the space is, alas, less nice.
In the finite versions of Toom's two-dimensional non-ergodic models as well
as of the complex one-dimensional models~\cite{Gacs1dim86,GacsSorg01},
the lower bound on
relaxation time \emph{grows} exponentially with the system size \( L \) (number of cells)
while here it decreases as \( L^{-1} \).

The papers~\cite{ToomGrowthI94,ToomGrowthII94} cited above can also be seen
as surface growth models where despite the fact that noise drives the surface to grow exclusively
in the upward direction, with probability 1 the surface height remains bounded.
The physics literature calls this \df{pinning}.
In~\cite{ToomPinning98}, Toom returned to this problem in a one-dimensional model in
which cells can have real number states.
The local rule drives the state in each step towards the local minimum of the three neighbors,
with a speed \( 1-\alpha \), but again there is a small rate \( \beta \) of random growth added.
The result is that this system has a property he calls \df{pseudo-pinning}: there is growth, but its
velocity is bounded \( C_{1}\alpha^{C_{2}/\beta} \), so it decreases exponentially in \( 1/\beta \).
(There is a somewhat similar lower bound.)
The proofs use multiple times a technique involving a majorizing operation \( \prec \)
in the following way:  \( AB\prec B'A' \) where operators \( A,B \) are replaced with operators
\( A',B' \) having slightly different parameters.

\section{Disappearing cells}

It is a natural question to try to generalize the cellular automata model by allowing the birth and
death of cells---in the sense that for example in one dimension
when a cell dies its site disappears, and its nearest neighbors connect with each other.
For a finite set of sites there are natural ways to do this in general form
even for random evolutions, see
Malyshev's~\cite{MalyshevRandomGraphs98}.
However, it does not seem possible to talk about, for example,
an infinite two-dimensional cellular automaton
in which cells (sites) can be eliminated or added: we would have to jump to the case of general
infinite graphs.

The one-dimensional case, the subject of
Toom's pioneering~\cite{ToomVarLength02} is an exception, with more technical
results in~\cite{ToomVarLength04}.
The model is attractive as it promises a new kind of
simple one-dimensional noisy ``non-ergodicity''.
Recall that with a binary set of states \( \states=\setof{0,1} \),
we may want to remember, in low-level noise, whether we started from all 0's or all 1's.
When in a sea of 0's a large island of 1's appears, we did not find any
\emph{simple} local rule capable of erasing it (in noise).
But when cells can be removed there is such a rule: for example just remove
every pair of cells containing 01!
Repeated application of this rule would eat up the island---hopefully even in noise.

Right at the start, however, Toom is faced with a new kind of problem:
how to define the probability space in question.
If a cell is deleted then the position of cells over a whole half-line changes.
The model he offers does not define a probability measure over histories,
only considers measures over the space of configurations.
Even here it is restricted to translation-invariant measures, so the history
is just a sequence of such measures \( \mu_{0},\mu_{1},\dots \).
The main definition is that of a transition operator \( P \) with \( \mu_{t+1}=P\mu_{t} \).
It is more complex than in the ordinary case, and
the operator obtained is \emph{not linear}.
The techniques relying on linearity are not available, so even a new proof of the
existence of an invariant measure is needed:
Toom provides this in~\cite{ToomContinuous07}, relying on continuity and the appropriate
fix-point theorems in functional analysis.
A generalization allowing arbitrary local (one-dimensional)
substitutions appeared in the joint paper~\cite{RochaSimasToomSubst11}
with Toom's students Rocha and Simas.
Even more detail can be found in the book-length exposition~\cite{ToomRamosRochaSimas11}
of Toom with his students.

In~\cite{ToomVarLength04}, Toom proves non-ergodicity of his variable-length medium
only for one-sided noise.
It is plausible that it would also hold for arbitrary small (independent)
noise; Toom promised but did not live to deliver this result.
The proof in the paper is a very detailed contour argument, in its general shape
not unlike the one Toom introduced for the Stavskaya model in~\cite{ToomOnStavskayaEng68}.

\section{The life}

(I posted an English translation of Toom's autobiographical notes in~\cite{ToomAutobio05}.)
Toom's activities extended far beyond research in mathematics.
In the latter he focused indeed mainly on probabilistic
cellular automata, 
but to computer scientists his name is probably best known for 
the work he did as an undergraduate.
Following the surprising discovery of Karatsuba that two \( n \)-digit
numbers can be multiplied in \( O(n^{\log_{2}3}) \) bit-operations (instead of the \( O(n^{2}) \)
steps of the thousand-years old school algorithm), he published in~\cite{ToomMult63} an algorithm
doing this in \( O(n^{1+\eps}) \) steps.
Stephen Cook found a similar algorithm independently
at about the same time.

\begin{sloppypar}
In the deep tradition of Russian mathematicians, among them many outstanding ones,
he was very active in mathematical
education---considering an honor to be able to contribute to the system he himself benefited from.
We can get an impression of the extent of his activity from the corresponding parts of his
homepage, now lost but mostly saved in \url{https://cs-web.bu.edu/faculty/gacs/toomandre-com-backup}:
\url{/my-articles/ruseduc}, \url{/my-articles/engeduc}.
He wrote a number of articles in the magazine Kvant
(addressed to interested and able high-school students), posing problems and giving little
expositions.
He was also a main organizer of and contributor to the so-called
``School by correspondence''.
This institution addressed the need of motivated students who wanted to go beyond
what their schools could offer but, living far from metropolitan areas (before the era of internet)
had no access to other opportunities.
They received challenging mathematical problems, had some time (say, a month) to work on them
and send in the solutions.
Professional mathematicians like Toom sent back the commented solutions, even giving
the students a second chance to solve the problem correctly.
In Moscow, Toom ran a popular computer club, which also turned out to be a greatly successful way of
leading a number of youngsters towards professional mathematics and computer science.
\end{sloppypar}

In 1989, Toom moved to the United States;
among all the new social and existential challenges he had to face,
his interest and activism in education never diminished.
Ideally, given the depth of his experience and commitment, he should have thrived here,
but over eight years of trying, he did not secure a tenured
position in the United States.
Paradoxically, his passionate educational interest worked against him, because of one
handicap: a complete lack of the diplomatic gene.
He was successful in the classroom, but his rather bluntly expressed
opinions, starting with~\cite{ToomRussianTeacher93}, made him unattractive to search
committees.
In Russia, he did criticize some buerocratic aspects of the educational system, but
in America he mostly posed the Russian way of mathematics teaching as a superior example
and attacked the core institutional principles, as expressed for example in some standards documents,
of the American system (and the Brazilian one trying to follow it).
One of his favorite topics was ``word problems''; he argued their usefulness in great,
inspiring and convincing detail (and also practiced in the classroom what he preached).
Thanks to the respect he commanded in the strong probability-theory community of Brazil 
he ended up there (learning Portuguese at an advanced age).
Settling at the Federal University of Pernambuco  in Recife
he felt finally able to teach what he wanted and the way he wanted.

Andrei never considered himself merely a mathematician, and has always tried to
apply his rigorous thinking to other areas.
In the early 1970's he was drawn into the circle
of the psychologist Vladimir Lefebvre (later a professor at UC Irvine), who introduced
a formal, algebraic system of social interactions.
Toom wrote some papers on the subject, for example applying those concepts to
game theory.
But more importantly, he took up the idea of ``reflexion'' as a useful tool of analyzing
human behavior and interpreting literature.
An interesting example is his study of the famous---and rather enigmatic---novel
``A hero of our time'' by the 19th century poet and writer Mikhail Lermontov.
In his retirement in New York, the work cut short by his death
was curating, together with his wife Anna, the rich legacy 
of his grandfather, a noted Russian poet Antokolsky; they published some of
their new discoveries in~\cite{ToomsAntokolskyTsvetaeva22}.

\bibliographystyle{imsart-nameyear} 
\bibliography{reli,gacs-publ}


\end{document}